\documentclass[floatfix,eqsecnum,aps,nofootinbib, amssymb]{revtex4}

\usepackage{graphicx}
\usepackage{dcolumn}
\usepackage{bm}
\usepackage{graphics}
\usepackage{amsmath}
\usepackage{amssymb}
\usepackage{amscd}
\usepackage{afterpage}
\usepackage{float,times}
\usepackage{subfigure}
\usepackage{rotating}
\usepackage{multirow}
\usepackage{fancyheadings}
\usepackage{epsfig}
\usepackage{theorem}
\usepackage{moreverb}
\usepackage{euscript}
\usepackage{psfrag}

\begin{document}
\title{Quartification On An Orbifold}
\author{Alison Demaria}\email{a.demaria@physics.unimelb.edu.au} \affiliation{School
of Physics, Research Centre for High Energy Physics, The University of
Melbourne, Victoria 3010, Australia} \author{Kristian L. McDonald}\email{k.mcdonald@physics.unimelb.edu.au} \affiliation{School
of Physics, Research Centre for High Energy Physics, The University of
Melbourne, Victoria 3010, Australia} 

\begin{abstract}
We investigate quartification models in five dimensions, with the fifth
dimension forming an $S^1/Z_2\times Z_2'$ orbifold. The orbifold
construction is combined with a
boundary Higgs sector to break the quartified gauge group directly to
a group
$H\subset SU(3)^4$ which is operative at the electroweak scale. We
consider $H=G_{SM}\otimes SU(2)_\ell$  and $H=G_{SM}$, where $G_{SM}$ is the
standard model gauge group, and find that unification occurs only when
 the remnant leptonic colour symmetry $SU(2)_\ell$ remains
 unbroken. Furthermore, the demands of a realistic low energy fermion
 spectrum specify a unique symmetry breaking route for the unifying
 case of
 $H=G_{SM}\otimes SU(2)_\ell$. We contrast this with four
 dimensional quartification models where unification may be achieved
 via a number of different symmetry breaking routes both with and
 without the remnant $SU(2)_\ell$ symmetry. The boundary Higgs sector of our model
 may be decoupled to achieve a Higgsless limit and we show that the 
electroweak Higgs doublet may be identified as the
 fifth component of a higher dimensional gauge field.

\end{abstract}

\maketitle

\section{introduction}
The notion of quartification is predicated upon the hypothesis of
quark-lepton universality at high energies. Quark-lepton universality
is typically implemented by the imposition of a discrete interchange
symmetry between quarks and leptons, a demand which further requires
the introduction of a leptonic colour gauge group $SU(3)_\ell$. This
leads one to the so called quark-lepton symmetric
model~\cite{Foot:1990dw,Foot:1990um,Foot:1990un,Foot:1991fk,Levin:1993sq,Shaw:1994zs,Foot:1995xx,McDonald:2006dy}
wherein the known leptons are identified as one member of a generalised
$SU(3)_\ell$ lepton triplet.

One of the primary goals of modern theoretical
particle physics is to identify extensions to the standard
model (SM), with some of the most appealing extensions being grand
unified theories (GUTs). Adding a leptonic colour group to the SM
clearly renders the traditional approaches to unification,
namely $SU(5)$ and $SO(10)$, inapplicable. It was the desire to
uncover a unified theory capable of accommodating the notion of quark-lepton
universality which motivated the construction of the quartification
model~\cite{Joshi:1991yn,Babu:2003nw}. Borrowing from the notion of
trinification~\cite{Glashow:1984gc,Babu:1985gi}, which
postulates the gauge group $G_T=SU(3)_c\otimes SU(3)_L\otimes SU(3)_R$
augmented by a cyclic $Z_3$ symmetry, the
quartification model posits the gauge group $SU(3)^4\equiv G_T\otimes
SU(3)_\ell$ with a cyclic $Z_4$ symmetry permuting the group
factors. The discrete groups ensure a single coupling constant at the unification scale 
and thus, a crucial feature of GUTs, namely the unification of the gauge coupling 
parameters, can ensue. 
Whilst in its original implementation only partial
unification was achieved~\cite{Joshi:1991yn}, it was subsequently demonstrated that
the modification of the exotic lepton mass spectrum permitted full 
unification~\cite{Babu:2003nw}. This motivated a recent study of quartification
models with intermediate symmetry breaking scales~\cite{Demaria:2006uu,Demaria:2005gk} where it
was shown that unification could be achieved via a number of different
symmetry breaking routes.
 
In~\cite{Demaria:2006uu,Demaria:2005gk} a subset of schemes which allowed 
for unification and maintained
phenomenological consistency were uncovered.
Although representing an improvement on the original quartified theories of 
Refs.~\cite{Joshi:1991yn} and \cite{Babu:2003nw}, these models exhibited some of the unsatisfactory
features common to conventional GUTs, resulting mostly from the Higgs sector. 
A large number of scalar multiplets were required to achieve a realistic
model and any potential describing their self-interactions would contain 
a plethora of unknown parameters. Furthermore, the vacuum expectation
values (VEVs) and mass spectra of the fields was not predicated 
by the theory, forcing a variety of assumptions on this sector. 

Orbifold compactifications, however, uncover a new arena by which to
explore GUTs.
The compactification process provides a geometrical origin for the 
symmetry breaking, allowing for the removal of the scalar sectors 
which complicate conventional 4d unified theories. 
The ability to obtain a consistent effective field theory on these 
constructions has motivated the use of orbifolds to probe grand 
unified theories~\cite{orbifolds,gauge3}. 
Even the SM Higgs field has a natural origin in orbifold models, 
with gauge-Higgs unified theories identifying the Higgs doublet as higher dimensional components of the gauge fields \cite{gaugehiggs}. 

This type of hybrid model was recently explored in a supersymmetric trinification theory by Carone and 
Conroy~\cite{Carone:2004rp,Carone:2005rk}. They placed the trinification gauge supermultiplets on 
a five dimensional orbifold, and localised the full trinified matter content onto a brane. The orbifold action reduced 
the symmetry down to a subgroup on this subspace. This symmetry was then broken to the SM gauge group by a boundary scalar sector which 
decoupled from the low-energy theory. Additionally, the SM Higgs doublets were recovered as 
remnant zero modes of the gauge fields.  

The realisation of a GUT that is not reliant upon a scalar sector and its associated problems provides the 
motivation to explore quartification in this orbifold context. 
We follow a similar approach to Carone and Conroy~\cite{Carone:2005rk}, 
considering a five dimensional quartification model and studying the
prospects for unification within this framework. Interestingly we find
that unification is only achieved when the remnant leptonic colour
group $SU(2)_\ell\subset SU(3)_\ell$ remains unbroken. Furthermore the
demands of a phenomenologically acceptable fermion spectrum specifies
the choice of orbifold symmetry reduction on the SM matter brane and
thus we arrive at a unique five dimensional quartification model which
achieves unification.

The layout of the paper is as follows. Section~\ref{sec:quart_prelim}
provides the reader with a brief summary of the quartification framework. In
Section~\ref{sec:quart_symmetry_b} we consider the higher dimensional
symmetry breaking necessary to reduce $SU(3)^4$ to an acceptable low
energy subgroup. The symmetry breaking occurs in two distinct ways, via
both orbifolding and the introduction of a boundary Higgs sector (which
we ultimately decouple from the theory), and we
detail each of these mechanisms. In
Section~\ref{sec:quart_fermion_mass} we consider fermion mass in the
model and demonstrate that the exotic fermions
all obtain GUT scale masses whilst the SM fermions remain massless
until electroweak symmetry breaking
occurs. Section~\ref{sec:quart_unification} covers the issue of gauge
coupling
unification within the model and we contrast our five dimensional model
with conventional approaches in
Section~\ref{sec:quart_comparison}. We conclude in
Section~\ref{sec:quart_conc}.
\section{Quartification Preliminaries\label{sec:quart_prelim}}
In this section we briefly surmise the conventional quartification
framework. 
For more details refer to 
~\cite{Babu:2003nw, Demaria:2005gk, Demaria:2006uu}. The quartification gauge group is 
\begin{equation}
\centering
G_4 = SU(3)_q \otimes SU(3)_L \otimes SU(3)_{\ell} \otimes SU(3)_R.
\label{eqn:quart}
\end{equation} 
A $Z_4$ symmetry which cyclicly permutes the gauge groups as per $q \rightarrow L \rightarrow 
\ell \rightarrow R \rightarrow q$ is also imposed to ensure a single gauge
coupling constant. 
The fermion assignment is anomaly-free, with each family contained within a left-handed ${\bf 36}$ 
of Eq.~\ref{eqn:quart}
\begin{eqnarray}
\centering
{\bf 36}& =& ({\bf 1,1,3, \overline{3}}) \oplus ({\bf {\overline{3}},1,1,3}) \oplus ({\bf 3 ,\overline{3},1,1}) \oplus 
( {\bf 1, 3 ,\overline{3},1}) \\ \nonumber
& = & \qquad \ell^c \qquad \!\! \oplus \qquad q^c \qquad \!\!\! \oplus \qquad q \qquad \oplus \qquad \ell.
\end{eqnarray}
$q$ denotes the left-handed quarks, $\ell$ the left-handed leptons and $q^c$ and $\ell^c$
the left-handed anti-quarks and anti-leptons respectively. These are represented by $3 \times 3$ 
matrices, with the first family having the form 
\begin{eqnarray}
\centering
q & \sim & \left( \begin{array}{ccc} 
d & u & h \\
d & u & h \\
d & u & h \end{array} \right), \qquad q^c \sim \left( \begin{array}{ccc}
d^c & d^c & d^c \\
u^c & u^c & u^c \\
h^c & h^c & h^c \end{array} \right),  \nonumber \\
\ell & \sim & \left( \begin{array}{ccc} 
x_1 & x_2 & \nu \\
y_1 & y_2 & e \\
z_1 & z_2 & N \end{array} \right), \qquad \ell^c \sim  \left( \begin{array}{ccc}
x_1^c & y_1^c & z_1^c \\
x_2^c & y_2^c & z_2^c \\
\nu^c & e^c & N^c \end{array} \right). \label{eq:fermions}
\end{eqnarray}
The quark multiplets contain exotic quark colour triplets $h$ and $h^c$, and exotic fermions 
are also required to fill the lepton representations. The SM leptons have partners 
$x_i,\,x_i^c, \, y_i,\,y_i^c, \,z_i$ and $z_i^c, \, i=1,2$ which transform as $SU(2)_\ell$ 
doublets, and there are two exotic singlets $N$ and $N^c$. 
Under $H=G_{SM}\otimes SU(2)_\ell$, the generator of hypercharge is 
\begin{equation}
\centering
Y=\frac{1}{\sqrt{3}} \left( T_{8L} + T_{8\ell}+ T_{8R} \right) + T_{3R}
\end{equation}
where the $T$'s refer to the Gell-Mann generators. In this framework, the exotic particles 
have the electric charges $Q(x_i,y_i,z_i)=(1/2,-1/2,1/2)$, $Q(N)=0$, and $Q(h)=-2/3$. 
In the case where the leptonic colour symmetry is entirely broken $H=G_{SM}$, the 
$x_1,x_1^c,y_2, y_2^c, z_1$ and $z_1^c$ leptons become neutral. 

The Higgs sector required to reproduce acceptable low energy
phenomenology in 4d quartification models \cite{Babu:2003nw, Demaria:2005gk,Demaria:2006uu} 
consists of two different
$\mathbf{36}$'s of $G_4$ which are labelled as per
\begin{eqnarray}
\centering
\Phi_a & \sim (\mathbf{1,3,1, \overline{3}}), \qquad 
\Phi_b \sim (\mathbf{\overline{3},1,3,1}), \qquad 
\Phi_c  \sim (\mathbf{1, \overline{3},1,3}), \qquad  
\Phi_d \sim (\mathbf{3,1, \overline{3},1}), \nonumber \\
 \Phi_\ell & \sim (\mathbf{1, 3, \overline{3},1}), \qquad 
\Phi_{\ell^c} \sim (\mathbf{1, 1,3, \overline{3}}), \qquad 
\Phi_{q^c} \sim (\mathbf{\overline{3},1,1,3}), \qquad 
\Phi_{q} \sim (\mathbf{3, \overline{3},1,1}).
\label{eqn:higgstrans}
\end{eqnarray}
These fields are closed under the $Z_4$ symmetry and generate 
realistic fermion masses and mixings.
We shall not require the two $\mathbf{36}$'s of scalars to accomplish
the necessary symmetry breaking in the 5d model, relying instead on the orbifold geometry. 
The subsequent absence of scalars with non-trivial $SU(3)_q$ quantum numbers removes issues of 
proton stability in this higher dimensional context. 

\section{Gauge Symmetry Breaking Framework\label{sec:quart_symmetry_b}}
\subsection{Orbifold Symmetry Breaking}
We consider a pure quartification gauge theory defined in a five dimensional spacetime, with the extra dimensional 
coordinate labelled $y$. This extra dimension is compactified on an $S^1/Z_2 \times Z_2'$ orbifold. This 
procedure constrains the spacetime geometry as per
\begin{equation}
\centering
y  \rightarrow  y + 2 \pi R  , \qquad y  \rightarrow  -y, \qquad
y'  \rightarrow  -y', 
\end{equation}
where $y' \equiv y + \pi R /2$. 
These identifications have the effect of reducing the physical region to the 
interval $y \in [0, \pi R/2]$ with two fixed points located at $y=0$ and $y=\pi R/2$. 
These points are geometrical singularities and are thus chosen as the location 
of 4d branes. 
For simplicity, 
we shall consider 
all matter fields to be localised on the $y=\pi R/2$ brane, which we shall refer to as the matter 
brane, while the 
brane at $y=0$ remains ``hidden''. 

The orbifold action also has a definition on the space of gauge fields which freely propagate 
in the bulk. 
The quartification gauge fields are denoted by 
\begin{eqnarray}
A^M(x^{\mu},y)&=&A^M_q (x^{\mu},y)\oplus A^M_L(x^{\mu},y)\oplus A^M_{\ell}(x^{\mu},y)\oplus A^M_R(x^{\mu},y), \\
&=& A_q^{M\,a} \,T^a \oplus A_L^{M\,a}\, T^a\oplus A_{\ell}^{M\,a}\, T^a\oplus A_R^{M\,a} \,T^a 
\end{eqnarray}
where $a=1,2,...,8$ is the gauge index 
and $M$ is the Lorentz index $M=\mu,5$. 
We define $P$ and $P'$ to be $3 \times 3$ matrix 
representations of the orbifold actions $Z_2$ and $Z_2'$ respectively. To maintain gauge invariance under these projections, 
the gauge fields are required to have the transformations
\begin{eqnarray}
A_{\mu}(x^{\mu},y) & \rightarrow & A_{\mu}(x^{\mu},-y) = P A_{\mu}(x^{\mu},y) P^{-1},\\ \nonumber
A_{5}(x^{\mu},y) & \rightarrow & A_{5}(x^{\mu},-y) = - P A_{5}(x^{\mu},y) P^{-1}, \\ \nonumber
A_{\mu}(x^{\mu},y') & \rightarrow & A_{\mu}(x^{\mu},-y') = P' A_{\mu}(x^{\mu},y') P'^{-1}, \\ \nonumber
A_{5}(x^{\mu},y') & \rightarrow & A_{5}(x^{\mu},-y') = - P' A_{5}(x^{\mu},y') P'^{-1}.\label{eq:projections}
\end{eqnarray}
Given that $P$ and $P'$ define a representation of reflection symmetries, their eigenvalues are $\pm 1$, and thus we can 
express these matrices in diagonal form, with a freedom in the parity choice of the entries. 
The exact nature of these actions then completely determines the gauge symmetry which remains unbroken at the fixed points. 
Unless $P$ is the identity matrix, not all the gauge fields will commute 
with the orbifold action. 
These fields are projected off the brane, and thus only a subset of the 
5d gauge theory is manifest at the fixed points. 
Ideally, one would desire the matter 
brane to respect only the SM gauge group; however, this is not directly possible here via orbifolding. 
The $Z_2 \times Z_2'$ actions are abelian and commute with the diagonal quartification generators. 
Subsequently, the rank of the quartification group must be preserved on the matter brane. This means that breaking unwanted 
$SU(3)$ factors has the trade-off of retaining the spurious $U(1)$
subgroups. Thus one must invoke a mechanism in tandem to 
orbifolding in order to accomplish the breaking to $G_{SM}$. 

We shall choose to have Higgs fields $\chi_i$ localised on $y=\pi R/2$ to instigate the rank-reducing breaking, 
giving the general symmetry breaking framework 
\begin{equation}
F \equiv SU(3)^4 \stackrel{\textrm{orbifold}}{\longrightarrow} G \stackrel{\textrm{Higgs}}{\longrightarrow} H \equiv G_{SM} \otimes SU(2)_\ell.
\end{equation}
This type of hybrid model has been explored recently in the context of
$SO(10)$ \cite{Kim:2002im} and the aforementioned $SU(3)^3$~\cite{Carone:2005rk, Kim:2004pe} 
orbifold GUTs. 
We shall consider the case where there is a residual $SU(2)_\ell$
symmetry in what follows. It is
phenomenologically acceptable to retain $SU(2)_\ell$ as an exact low
energy symmetry. This symmetry acts only on exotic leptons which
fill out the $SU(3)_\ell$ multiplets (known as liptons in the literature). These are constrained to be
heavier than a TeV and will be much heavier than this in our
construct. We shall comment on the case when $SU(2)_\ell$ is broken
(namely $H=G_{SM}$) in
Section~\ref{sec:quart_broken_su2}.

The group $G$ is determined by the desire to reproduce the correct SM matter content on the brane at the $y=\pi R/2$ 
fixed point. It turns out that the 
only feasible choice is to consider the orbifold returning the symmetry 
$G_{\pi R/2} \equiv SU(3)_q \otimes SU(2)_L \otimes SU(3)_\ell \otimes SU(3)_R \otimes U(1)_L$. 
All other options do not return a favourable low energy theory. For example, we require the 
$SU(3)_R$ symmetry to generate realistic quark masses, and the $SU(2)_L$ symmetry must be 
realised so as to return a SM Higgs. The full leptonic colour symmetry $SU(3)_\ell$ also 
needs to be respected on the matter brane. If this symmetry was broken, then 
indistinguishability issues surface between the SM leptons and their liptonic partners.

The orbifold action can be decomposed as 
$(P,P') = (P_q \oplus P_L \oplus P_{\ell} \oplus P_R,\,P_q' \oplus P_L' \oplus P_{\ell}' \oplus P_R')$, where 
\begin{eqnarray}
P_q & = & \textrm{diag}(1,1,1), \;\;\;\; P_L = \textrm{diag}(1,1,-1), \;\;\;\;
P_{\ell} = \textrm{diag}(1,1,-1), \;\;\;\;\; P_R = \textrm{diag}(1,1,-1),  \nonumber \\
P_q' & = & \textrm{diag}(1,1,1), \;\;\;\; P_L' = \textrm{diag}(1,1,-1), \;\;\;\; 
P_{\ell}' = \textrm{diag}(1,1,1), \qquad P_R' = \textrm{diag}(1,1,1).\label{eq:proj1}
\end{eqnarray}
Under this action, the parity assignments of the vector components of the gauge multiplets are  
\begin{eqnarray}
A_q^{\mu} &=& \left( \begin{array}{ccc} 
(+ \, ,+ ) &  (+ \, ,+ )& (+ \, ,+ )\\
(+ \, ,+ ) & (+ \, ,+ ) & (+ \, ,+ )\\
(+ \, ,+ ) & (+ \, ,+ ) & (+ \, ,+ ) \end{array} \right), \qquad 
A_L^{\mu} = \left( \begin{array}{ccc} 
(+ \,, + ) & (+ \, ,+ )& (- \, ,-) \\
(+ \, ,+ ) & (+ \, ,+ ) & (- \, ,- ) \\
(- \,, -) & (- \, ,-) & (+ \, ,+ )\end{array} \right), \\ \nonumber
A_{\ell}^{\mu} & = & \left( \begin{array}{ccc}
(+ \, ,+ ) & (+ \, ,+ )& (- \, ,+) \\
(+ \, ,+ ) & (+ \, ,+ ) & (- \, ,+ ) \\
(- \, ,+) & (- \, ,+) & (+ \, ,+ )\end{array} \right), \qquad
A_R^{\mu} = \left( \begin{array}{ccc}
(+ \, ,+ ) & (+ \, ,+ )& (- \, ,+ )\\
(+ \, ,+ )& (+ \, ,+ )& (- \, ,+ )\\
(- \, ,+ )&(- \, ,+ )&(+ \, ,+ ) \end{array} \right).
\end{eqnarray}
The wavefunctions of these component fields have the Kaluza-Klein decompositions 
\begin{eqnarray}\label{eq:modes}
\centering
A^\mu_{++}(x^{\nu},y) &=& \frac{1}{\sqrt{2\,\pi R}}A^{\mu \, (0)}_{++}(x^{\nu})+\frac{1}{\sqrt{\pi R}} \sum^{\infty}_{n=1} A^{\mu \, (2n)}_{++}(x^{\nu}) \, \cos \left(\frac{2ny}{R}\right), \\
A^\mu_{-+}(x^{\nu},y) &=& \frac{1}{\sqrt{\pi R}} \sum^{\infty}_{n=0} A^{\mu \, (2n+1)}_{-+}(x^{\nu}) \, \sin \left(\frac{(2n+1)y}{R}\right), \\
A^\mu_{+-}(x^{\nu},y) &=& \frac{1}{\sqrt{\pi R}} \sum^{\infty}_{n=0} A^{\mu \, (2n+1)}_{+-}(x^{\nu}) \, \cos \left(\frac{(2n+1)y}{R}\right), \\
A^\mu_{--}(x^{\nu},y) &=& \frac{1}{\sqrt{\pi R}}\sum^{\infty}_{n=0} A^{\mu \, (2n+2)}_{--}(x^{\nu}) \, \sin \left(\frac{(2n+2)y}{R}\right), 
\end{eqnarray}
The towers of $4d$ fields $A^{\mu \, (2n)}_{++}(x^\nu),\, A^{\mu \, (2n+1)}_{-+}(x^\nu),\, A^{\mu \, (2n+1)}_{+-}(x^\nu),\,A^{\mu \, (2n+2)}_{--}(x^\nu)$ 
upon compactification acquire KK masses $2 nM_c,\,(2 n+1)M_c, \,(2 n+1)M_c,\,(2 n+2)M_c$ respectively where $M_c \equiv 1/R$ 
is defined to be the compactification scale. 
Only the $(+,+)$ fields possess massless zero modes, and hence the massless sector is restricted to a subset of the full 5d theory. 
The towers $(-,+)$ and $(+,-)$ have degenerate mass spectra, as do the towers $(-,-)$ and $(+,+)$ excluding the zero mode. 

We can now discern the nature of the theory on the branes. 
In general, only those fields which are odd under $P\,(P')$ vanish at $y=0 \,(y'=0)$. 
Thus, one can see that at $y=0$, the gauge symmetry $SU(3)_q \otimes SU(2)_L \otimes SU(2)_{\ell} \otimes SU(2)_R \otimes U(1)^3$ 
is respected, while at the fixed point $y=\pi R/2$ the symmetry is 
$SU(3)_q \otimes SU(2)_L \otimes SU(3)_{\ell} \otimes SU(3)_R \otimes U(1)_L$.    
The cumulative effect of the orbifold compactification 
is the reduction of the overall symmetry to the intersection of these two subgroups, 
$G\equiv G_{y=0} \cap G_{y=\pi R/2} = SU(3)_q \otimes SU(2)_L \otimes
SU(2)_{\ell} \otimes SU(2)_R \otimes U(1)^3$. Clearly further symmetry
breaking is required to firstly reproduce, then break, the SM.

As evident in Eq.~\ref{eq:projections}, the scalar components $A^5_i$ necessarily have opposite parities to
their vector counterparts allowing one to determine the scalar content at the fixed points. 
The transformation of $A_L^{5\,a}$ importantly reveals a scalar
doublet which has a massless zero mode at the 4d level, 
and this possesses the appropriate hypercharge to be identified as a SM Higgs doublet.
This type of identification is referred to as gauge-Higgs unification and has been employed extensively 
as an origin for the EW Higgs in orbifold GUTs~\cite{gaugehiggs}. Attention to this sector will be given in Section~\ref{cha:EWhiggs}.  

\subsection{Brane Breaking}\label{cha:branebreaking}
After the orbifold compactification, the symmetry of the 4d effective theory is 
$SU(3)_q \otimes SU(2)_L \otimes SU(2)_\ell \otimes SU(2)_R \otimes U(1)^3$, while on the matter 
brane the gauge symmetry is 
$G_{\pi R/2} \equiv SU(3)_q \otimes SU(2)_L \otimes SU(3)_\ell \otimes SU(3)_R \otimes U(1)_L. $
The usual approach to breaking this remaining symmetry is to implement 
boundary conditions on the compactified space 
for the gauge fields \cite{Csaki:2003dt}. 
This method is more generalised than the orbifold mechanism, leading to a greater set of 
symmetry breaking opportunities. 

The structure of the boundary conditions are realised in a UV
completion of our GUT theory in which localised
scalar fields $\chi_i$ prescribe the breaking. As the fields are localised on the matter brane where only the 
reduced $G_{\pi R/2}$ symmetry is operative, the fields are housed in incomplete quartification multiplets, 
with transformation properties 
\begin{equation}
\chi_{1,2} \sim (\mathbf{1,1,1,3}), \qquad \chi_{3} \sim (\mathbf{1,1,3,1}), \qquad \chi_4 \sim (\mathbf{1,1,3,\overline{3}}).
\end{equation}
The three fields $\chi_{1,2,3}$ carry $U(1)_L$ charge $-2/\sqrt{3}$
and are necessary for the 
$SU(2)_R \otimes U(1)^3 \rightarrow U(1)_Y$ breaking. 
The final Higgs field, $\chi_4$ is neutral under $U(1)_L$. Its VEVs do not increase the symmetry breaking 
but they are important contributors to fermion mass generation.  
As will be detailed later, we shall be considering the limit in which this sector decouples entirely from the 
brane, and so the addition of these scalars does not pose a complication. 

The effect of these fields can be ascertained from the kinetic sector of our action. The 5d action assumes the form 
\begin{equation}
\centering
\mathcal{S} \supset \int d^5x \left( -\frac{1}{4} F^a_{\mu \,\nu}\,F^{a\,\mu\,\nu}-\frac{1}{2}\,F^a_{5\,\nu}\,F^{a\,5\,\nu} 
+\left( D_\mu \, \chi_i^\dagger \, D^\mu \, \chi_i -V(\chi) \right)\delta(y-\pi R/2) \right),
\end{equation} 
and the relevant terms to consider are 
\begin{equation}
\centering
\int^{\pi R/2}_0 dy \left( -\frac{1}{2} \partial_5 \, A_{\mu}(x^\nu,y) \, \partial^5 \, A^{\mu}(x^\nu,y) + \frac{g_5^2}{4} \langle \chi \rangle^\dagger 
 \langle \chi \rangle \, A^{\mu}(x^\nu,y)\,A_{\mu}(x^\nu,y) \, \delta(y-\pi R/2) \right).
\end{equation}
$g_5$ is the five-dimensional gauge coupling constant defined by $g_5 \equiv g_4 \, \sqrt{\frac{\pi R}{2}}$ and has mass dimension $-1/2$. 
Variation in $A^{\mu}(x^\nu,y)$ gives, after integration over the extra dimension, the surface terms 
\begin{eqnarray}
\centering
&-\partial_5 \, A_\mu(x^\nu,y) \vert_{y=0}, \nonumber \\
&\left( -\partial_5 \, A_\mu(x^\nu,y)  + \frac{g_5^2}{2} \langle \chi \rangle^\dagger 
 \langle \chi \rangle \, A_{\mu}(x^\nu,y)\,\right) \, \delta A^{\mu}(x^\nu,y)\, \vert_{y=\pi R/2}
\end{eqnarray}
which must vanish.
Making the definition 
\begin{equation}\label{eq:Vdef}
\centering
V\equiv \frac{g_5^2}{2}\langle \chi \rangle^\dagger 
 \langle \chi \rangle,
\end{equation}
we obtain the boundary conditions 
\begin{eqnarray}
\centering
\partial_5 \, A_\mu(x^\nu,0)&=&0, \\
\partial_5 \, A_\mu(x^\nu,\pi R/2) &=& V\, A_\mu(x^\nu,\pi R/2),\label{eq:BC}
\end{eqnarray} 
which illustrate the constraints imposed on the gauge fields by the boundary Higgs sector. $V$, having dimensions of mass, 
reflects the order parameter of the symmetry breaking. If $V=0$, Eq.~\ref{eq:BC} reduces to Neumann boundary conditions 
and returns the usual orbifold behaviour. If $V \neq 0$, then the zero mode for $A_\mu$ is no longer massless, and one 
has gauge boson mass terms localised at the matter brane as we shall now clarify.  

These boundary conditions affect only the fields which are even at $y=\pi R/2$, with the towers corresponding 
to the parities $(\pm ,-)$ disappearing on the matter brane. The non-vanishing fields $A^{(+,+)}_\mu$ have the generic profile 
\begin{equation}
\centering
A_\mu(x^\nu,y) = N_n \, \cos({M_n y})\,A_\mu^{(n)}(x^\nu) 
\end{equation}
with KK mass $M_n$.
Enforcement of Eq.~\ref{eq:BC} appreciably modifies the wavefunctions of these gauge bosons, with their KK 
masses shifted by an amount governed by 
\begin{equation}\label{eq:constraint1}
\centering
M_n \tan({M_n \pi R/2})=-V. 
\end{equation}
For large values of $V$, Eq.~\ref{eq:constraint1} shows that the KK tower has a mass spectrum approximated by
\begin{equation}\label{eq:shift1}
\centering
M_n \approx M_c \,(2n+1) \left(1+\frac{M_c}{\pi V}+ \dots \right), \qquad n=0,1,2,...
\end{equation}
giving a tower with the lowest-lying states $M_c, \, 3M_c, \, 5M_c, ...$. This represents an offset of 
$M_c$ relative to the $V=0$ tower, with the field no longer retaining a massless zero mode. 
As $V \rightarrow \infty$, the brane localised mass terms for the gauge fields 
become larger and their wavefunctions are eventually expelled from the brane. 
A similar mass shift can be induced by the boundary conditions on the $(-,+)$ fields. 

From Eq.~\ref{eq:Vdef}, the association of $V$ with the VEVs of the boundary scalar 
sector implies that the limit $V \rightarrow \infty$ is attained when $\langle \chi \rangle \rightarrow \infty$.
However, when the VEVs of the Higgs fields are taken to infinity, the shift in the KK masses of the gauge fields is finite, 
giving the exotic gauge fields masses dependent only upon the compactification scale $M_c$. 
Consequently, these fields remain as ingredients in the effective theory while the boundary Higgs sector decouples entirely, and we 
can view our reduced symmetry theory in an effective Higgless limit. Interestingly, in this limit also, the high-energy 
behaviour of the massive gauge boson scattering remains unspoilt as
shown in~\cite{Csaki:2003dt}.

Depending on the exact nature of $V$, a shift can be induced in the KK towers of the gauge fields 
generating a greater symmetry breaking than the orbifold compactification. 
The gauge fields are decomposed into one of six possible KK towers depending on the orbifold 
and boundary conditions, and we need to choose the latter 
such that the $G_{\pi R/2}/H$ gauge fields are exiled from the brane, while those corresponding to the $H$ symmetry remain 
unperturbed. Given that the structure of the breaking parameter $V$ is predetermined by the Higgs fields, the desired 
boundary constraints can be satisfied with our scalar fields acquiring VEVs of the form
\begin{eqnarray}
\centering 
\langle \chi_1 \rangle &=& \left( \begin{array}{c}0 \\ 0 \\v_1 \end{array} \right), \qquad 
\langle \chi_2 \rangle = \left( \begin{array}{c}v_2 \\ 0 \\ v_3 \end{array} \right), \\
\langle \chi_3 \rangle &=& \left( \begin{array}{c}0 \\ 0 \\v_4 \end{array} \right), \qquad 
\langle \chi_4 \rangle = \left( \begin{array}{ccc}0 & 0 & 0 \\ 
0 & 0 & 0 \\ v_5 & 0 & v_6  \end{array} \right).
\end{eqnarray}
The fields $\chi_{1,2,3}$ and $\chi_4$ have a $U(1)_L$ charge of $-2/\sqrt{3}$ and $0$ respectively, 
where these assignments and the VEV structure are motivated by the embedding of the fields in the 
conventional quartification models~\cite{Demaria:2005gk}. 

These fields transform trivially under $SU(3)_q \otimes SU(2)_L$ and thus the gauge fields defining these symmetries remain 
unaffected by the scalars. The $SU(3)_\ell \otimes SU(3)_R$ gauge fields, however, are subject to the constraint 
\begin{equation}\label{eq:VBC}
\centering
\partial_5 A_\mu^i(x^\nu,\pi R/2) = V_{ij}A_\mu^j(x^\nu,\pi R/2)
\end{equation}
where $V_{ij}$ is a matrix in the space of these fields. For the non-abelian components 
this can be decomposed into the form 
\begin{eqnarray}
\centering 
\partial_5 A^{\mu \, a}_{\ell} (x^{\nu},\pi R/2) & =&0, \qquad \qquad \qquad \qquad \;\;\;\;a=1,2,3, \\
\partial_5 A^{\mu \,\hat{a}}_{\ell} (x^{\nu},\pi R/2)&=& V_{\ell} A^{\mu \,\hat{a}}_{\ell}(x^{\nu},\pi R/2),\qquad \;\hat{a}=4,5,6,7, \\
\partial_5 A^{\mu \,\hat{a}}_R (x^{\nu},\pi R/2)&=& V_R A^{\mu \,\hat{a}}_R(x^{\nu},\pi R/2), \qquad \hat{a}=1,2,4,5,6,7,
\end{eqnarray}
where the $a$'s refer to the unbroken group indices and $\hat{a}$ the broken indices. 
The $V_{\ell, R}$ entries have the general form $\sum_i c_i \,g_5^2 \,v_i^2$, for some constant $c_i$, 
with their precise values determined by the dynamics 
of the UV completed theory. 

The constraints on the $U(1)$ factors are more non-trivial. The hypercharge gauge field is 
\begin{equation}
\centering
A_Y^{\mu}=\left(\frac{2}{\sqrt{3}}\left( A^8_L + A^8_{\ell}+A^8_R \right)+2\,A_R^3 \right)^{\mu}
\end{equation}
and this must remain massless in the low-energy theory. Subsequently, this linear combination must 
correspond to the sole zero eigenvalue of the relevant subset of the generalised $V$ matrix of Eq.~\ref{eq:VBC}
defined by 
\begin{equation}
\centering
\partial_5 \hat{A}^{\mu}_i(x^{\nu},\pi R/2) =V_{ij}\hat{A}_j^{\mu}, \qquad i,j=1,2,3,4
\end{equation}
where 
\begin{equation}
\centering
\hat{A}^{\mu}\equiv \left( \begin{array}{cccc}A^{8\, \mu}_L(x^{\nu},\pi R/2) &A^{8\, \mu}_{\ell}(x^{\nu},\pi R/2)&A^{3\, \mu}_R(x^{\nu},\pi R/2)&A^{8\,\mu}_R(x^{\nu},\pi R/2) \end{array} \right)^T.
\end{equation}
The explicit form of $V_{ij}$ was determined in Ref.~\cite{Carone:2004rp} for a model described by the trinification gauge group, with the entries functions of $g_5^2v_i^2$ as defined by the VEV structure. 
This parametrisation generalises to the quartification case, and reveals a zero  
eigenvalue for the vector $1/\sqrt{3} \lbrace 1,1,\sqrt{3},1 \rbrace$, and non-zero eigenvalues for the 
remaining three eigenvectors, denoted as $A^\mu_{X_1},\,A^\mu_{X_2}$ and $A^\mu_{X_3}$. Consequently, the $(+,+)$ towers 
corresponding to these latter three physical fields have their mass spectra shifted by $M_c$ and they are expelled from 
the low-energy theory.

The effect of the implementation of these boundary conditions on the 32 quartification gauge fields is 
summarised in Table~\ref{the_one} where the fields have been listed in terms of their 
$G_{\pi R/2}$ representations. The mass spectra shown incorporates any induced shifts due to 
non-trivial values of $V$, from which it can be deciphered that only the vector fields respecting 
$G_{SM} \otimes SU(2)_{\ell}$ possess a massless mode, with the other $(+,+)$ towers misaligned. 
The new mass spectra of these shifted towers is now degenerate with the $(-,+)$ towers. 
Furthermore, the only scalar that has a massless mode 
transforms as an $SU(2)_L$ doublet (and its conjugate). This mode has a component that can not be gauged away, 
giving us a physical electroweak Higgs field. 

\begin{table}[h]
\begin{center}
\begin{tabular}{cccc}
\hline
\hline
\bf{Fields} & \bf{Representation} & \bf{KK tower} & \bf{Mass} \\ 
\hline
\hline
$A^{\mu \, a}_q, \,\,\, a=1,2, ...,8$ & $(\mathbf{8,1,1,1})$& $(+,+)$ & $ 2\,n M_c$ \\
$A^{5 \, a}_q, \,\,\, a=1,2, ...,8$ & $(\mathbf{8,1,1,1})$& $(-,-)$ & $2\,( n+1) M_c$ \\
\hline
$A^{\mu \, a}_L, \,\,\;\; a=1,2,3$ & $(\mathbf{1,3,1,1})$& $(+,+)$ & $ 2\,n M_c$ \\
$A^{5 \, a}_L, \,\,\, \;\;a=1,2,3$ & $(\mathbf{1,3,1,1})$& $(-,-)$ & $2\,(n+1) M_c$ \\
$A^{\mu \, \hat{a}}_L, \,\,\, \hat{a}=4,5,6,7$ & $(\mathbf{1,2,1,1})+ (\mathbf{1,2,1,1})$& $(-,-)$ & $2\,( n+1) M_c$ \\
$A^{5 \, \hat{a}}_L, \,\,\, \hat{a}=4,5,6,7$ & $(\mathbf{1,2,1,1})+(\mathbf{1,2,1,1})$& $(+,+)$ & $ 2\,n M_c$ \\
\hline
$A^{\mu \, a}_{\ell}, \,\,\, a=1,2,3$ & $(\mathbf{1,1,3,1})$& $(+,+, V=0)$ & $ 2\,n M_c$ \\
$A^{5 \, a}_{\ell}, \,\,\, a=1,2,3 $ & $(\mathbf{1,1,3,1})$& $(-,-)$ & $2\,( n+1) M_c$ \\
$A^{\mu \, \hat{a}}_{\ell}, \,\,\, \hat{a}=4,5,6,7$ & $(\mathbf{1,1,2,1})+ (\mathbf{1,1,2,1})$& $(+,-)$ & $(2 n+1) M_c$ \\
$A^{5 \, \hat{a}}_{\ell}, \,\,\, \hat{a}=4,5,6,7$ & $(\mathbf{1,1,2,1})+(\mathbf{1,1,2,1})$& $(-,+)$ & $( 2n+1) M_c$ \\
\hline
$A^{\mu \, a}_R, \,\,\, a=1,2$ & $\subset(\mathbf{1,1,1,3})$& $(+,+, V \rightarrow \infty)$ & $(2n+1) M_c$ \\
$A^{5 \, a}_R, \,\,\, a=1,2$ & $\subset(\mathbf{1,1,1,3})$& $(-,-)$ & $2\,(n+1) M_c$ \\
 $A^{\mu \, a}_R, \,\,\, a=4,5,6,7$ & $(\mathbf{1,1,1,2})+ (\mathbf{1,1,1,2})$& $(-,+)$ & $( 2n+1) M_c$ \\
$A^{5 \, a}_R, \,\,\, a=4,5,6,7$ & $(\mathbf{1,1,1,2})+(\mathbf{1,1,1,2})$& $(+,-)$ & $( 2n+1) M_c$ \\
\hline
$A^{\mu}_Y$ & $(\mathbf{1,1,1,1})$ & $(+,+,V=0)$ & $2\,nM_c$\\
$A^{\mu}_{X_1}$ & $(\mathbf{1,1,1,1})$ & $(+,+, V \rightarrow \infty)$ & $(2n+1) M_c$\\ 
$A^{\mu}_{X_2}$ & $(\mathbf{1,1,1,1})$ & $(+,+, V \rightarrow \infty)$ & $(2n+1) M_c$\\ 
$A^{\mu}_{X_3}$ & $(\mathbf{1,1,1,1})$ & $(+,+, V \rightarrow \infty)$ & $(2n+1) M_c$\\ 
\hline
\hline
\end{tabular}
\caption{The decomposition of the 32 gauge fields into their respective KK towers at the fixed point $y=\pi R/2$. 
Their representations are given in terms of the brane symmetry before the additional breaking via the localised Higgs 
sector. The towers $(\pm ,-)$ have wavefunctions which are odd on this brane, and thus are vanishing\label{the_one}.} 
\end{center}
\end{table}
To summarise, the orbifold compactification alone reduces the gauge symmetry on the two 
fixed points to 
\begin{eqnarray}
\centering
G_0&=&SU(3)_q \otimes SU(2)_L \otimes SU(2)_\ell \otimes SU(2)_R \otimes U(1)^3, \\
G_{\pi R/2}&=& SU(3)_q \otimes SU(2)_L \otimes SU(3)_\ell \otimes SU(3)_R \otimes U(1)_L,
\end{eqnarray}
which reduces the overall symmetry of the low-energy 4d theory to 
$SU(3)_q\otimes SU(2)_L \otimes SU(2)_\ell \otimes SU(2)_R \otimes U(1)^3$. 
The boundary scalar sector then instigates the further breaking 
\begin{equation}
\centering
G_{\pi R/2} \rightarrow  SU(3)_q \otimes SU(2)_L \otimes SU(2)_\ell \otimes U(1)_Y
\end{equation}
returning the desired $G_{SM} \otimes SU(2)_\ell$ symmetry operative at the EM level.

\section{Fermion Masses\label{sec:quart_fermion_mass}}
\subsection{GUT scale masses}
We now introduce matter into our scheme. 
Due to the identification of the Higgs field as a remnant zero mode arising 
from the extra dimensional component of the gauge fields, if the fermions were 
bulk fields, their coupling with the Higgs would be universal. 
Thus, it would prove difficult to recover the detailed structure of
Yukawa couplings in the low-energy theory. As an aside we note
that it may be interesting to study a higher dimensional
quartification model employing split fermions. The high degree of
symmetry present in the quartification model can be utilised to
motivate the fermion localisation pattern of a split fermion
model~\cite{Coulthurst:2006bc}, though presumably a different symmetry
breaking mechanism to that employed here would be required.
In this work we shall assume that the fermions are localised on the $y=\pi
R/2$ brane where
they are housed in representations that need only respect the 
$SU(3)_q \otimes SU(2)_L \otimes SU(3)_{\ell} \otimes SU(3)_R \otimes U(1)_L$ symmetry.

Although blind to the entire $SU(3)^4$ symmetry, we consider the full fermion content 
of minimal quartification so that we can achieve anomaly cancellation. The exotic content, it transpires, will 
become massive through couplings with the boundary Higgs sector. The fermion representations under the brane symmetry 
are defined as follows:
\begin{eqnarray}
\centering
q_1 &\sim& (\mathbf{3,2,1,1})\left(\frac{1}{\sqrt{3}}\right), \;\;\;\; q_2 \sim (\mathbf{3,1,1,1})\left(-\frac{2}{\sqrt{3}}\right), 
\;\;\; q^c \sim (\mathbf{\overline{3},1,1,3})(0), \\
\ell_1 &\sim& (\mathbf{1,2,\overline{3},1})\left(-\frac{1}{\sqrt{3}}\right), 
\;\;\; \ell_2 \sim (\mathbf{1,1,\overline{3},1})\left(\frac{2}{\sqrt{3}}\right), 
\;\;\;\;\; \ell^c \sim (\mathbf{1,1,3,\overline{3}})(0),
\end{eqnarray}
whose matrix representations can be deduced from Eq.~\ref{eq:fermions}. 

The boundary Higgs fields $\chi_i$ couple to the localised fermions giving 
six invariant Yukawa interaction terms 
\begin{eqnarray}\label{eq:orbYukawa}
\centering
& Y_1  \, q^c \, q_2\,  \overline{\chi}_1   \qquad 
 Y_2  \, q^c\, q_2 \, \overline{\chi}_2  ,\nonumber \\ 
& Y_3  \, \ell_1 \, \ell_1 \, \overline{\chi}_{3}, \qquad  
 Y_4  \, \ell_2 \, \ell^c \,\chi_1 , \qquad 
 Y_5  \, \ell_2 \, \ell^c\,\chi_2 , \qquad
 Y_6  \, \ell^c\, \ell^c \, \chi_{4}.
\end{eqnarray}
Eq.~\ref{eq:orbYukawa} gives the quark mass terms 
\begin{equation}\label{eq:qmass1}
\centering
(Y_1\,v_1+Y_2\,v_3)\,h^c\,h + Y_2 \, v_2 \, d^c \, h, 
\end{equation}
revealing mixing between the $h^c$ and $d^c$, giving only one linear combination that is massless. This we 
identify to be the physical left-handed $d$ antiquark. 
Similarly, the exotic leptonic components all mix.
These mixings are sufficient to generate GUT scale Dirac masses for all but one exotic lepton, leaving only 
the SM fields and the right-handed neutrino as the low-energy massless fermion spectrum. As these mass terms 
are proportional to $v_i$, in the decoupling limit $v_i \rightarrow \infty$, all the exotic fermions are removed from 
the low-energy theory along with the $\chi_i$ fields. 
\subsection{Electroweak Scale Masses And The Higgs}\label{cha:EWhiggs}
In addition to the fermion spectrum, we have an electroweak scalar doublet arising from the fields 
$A_L^{5\,\hat{a}}$ which has a zero mode on the brane. We denote this doublet as 
$\Phi\sim (\mathbf{1,2,1,1})(-1)$.
What remains now is to induce electroweak symmetry breaking and to generate the correct Yukawa structure of 
these fields with the fermions. However, given that the SM 
Higgs is a remnant of the higher dimensional gauge sector, 
its dynamics are still dictated by the bulk theory. 
Local couplings between $\Phi$ and the fermion fields are 
restricted by the higher dimensional gauge invariance. This can be seen explicitly 
by looking at the gauge transformation parameters. A general $SU(3)^4$ gauge transformation is defined by 
\begin{equation}
\centering
U = e^{i \, [ \eta(x^{\nu})\,\xi(y)]_j^{a}\,T_j^a}, \qquad j=q,L,\ell,R
\end{equation}
where we have assumed a separable form for the gauge parameters. 
At the fixed points the $y$-dependent parameters corresponding to the broken generators satisfy  
$\xi^{\hat{a}}_j = 0 $ and $ \partial_5 \,\xi_j^{\hat{a}}|_{y_f} \neq 0. $
A remnant of the bulk symmetry at $y=\pi R/2$, however, is the condition   
$A_i^{5\,\hat{a}} \rightarrow A_i^{5 \, \hat{a}} + \partial_5 \, \xi^{\hat{a}}_i. $
This shift is broken by any Yukawa couplings directly involving $\Phi$. 
As a result, the only gauge invariant interactions between the brane fermions and the SM Higgs field are non-local, 
involving Wilson lines. 

The Wilson line operator is defined as 
\begin{equation}
\centering
W = \mathcal{P} \, \exp \left(i \int_{y_i}^{y_f} \Phi \, dy \right)\vert_{\mathbf{R}}, 
\end{equation}
in the representation $\mathbf{R}$ 
where $\mathcal{P}$ is the path-ordered product, and $y_i$ and $y_f$ are fixed points of the orbifold. 
If $y_1=y_2=\pi R/2$, this operator 
is linear under the gauge transformations and thus can couple to the fermion fields. Its non-locality means that the Higgs 
potential generated at the quantum level is also non-local and thus insensitive to the UV physics. 
We assume that the generation of the Wilson lines is due to some
dynamics in the UV completion of the theory.
Irrespective of their origin, these operators are natural entities to consider, satisfying invariance under both the gauge symmetry and the 
orbifold projections.

The non-local Lagrangian has the general form 
\begin{equation}
\centering
\mathcal{L}^{NL}= \lambda \, f_1\vert_{y=y_i} \, W\, \chi\vert_{y=y_f} 
 \; + \;  \lambda' \, f_2\vert_{y=y_i} \, \overline{W} \, \overline{\chi}\vert_{y=y_f} + \, \dots
\end{equation}
where $f_1, \, f_2, \dots$ are localised fermion bilinears, and each term is 
suppressed by the appropriate power of the fundamental scale $\Lambda$. 
If the Higgs field acquires a non-trivial VEV, then its non-local interaction with the brane fields 
generates 
the effective 4d local Yukawa terms 
\begin{eqnarray}
\centering
& \frac{1}{\Lambda}\,\lambda_1 \, \overline{\chi}_1\,q_1\,q^c \Phi, \qquad 
\frac{1}{\Lambda}\lambda_2 \, \overline{\chi}_2\,q_1\,q^c \Phi, \qquad 
\frac{1}{\Lambda^2}\lambda_3 \, \chi_1 \, \chi_2 \,q_1\,q^c\,\overline{\Phi} \nonumber \\
& \frac{1}{\Lambda}\lambda_4 \, \overline{\chi}_4 \, \ell_1 \, \ell_2 \, \Phi, \qquad
 \frac{1}{\Lambda^2}\lambda_5 \, \overline{\chi}_1\,\overline{\chi}_4 \, \ell_1\,\ell_2 \,\Phi, \qquad 
 \frac{1}{\Lambda^2}\lambda_6 \, \overline{\chi}_2\,\overline{\chi}_4 \, \ell_1\,\ell_2 \,\Phi, \nonumber \\
& \frac{1}{\Lambda}\lambda_7 \, \chi_1 \, \ell_1 \, \ell^c \, \overline{\Phi}, \qquad 
\frac{1}{\Lambda}\lambda_8 \, \chi_2 \, \ell_1 \, \ell^c \, \overline{\Phi},\qquad
 \frac{1}{\Lambda^2}\lambda_9 \, \chi_3 \, \overline{\chi}_4 \, \ell_1\,\ell^c\,\overline{\Phi}, \qquad 
\frac{1}{\Lambda^2}\lambda_{10} \, \overline{\chi}_1 \, \overline{\chi}_2 \, \ell_1 \, \ell^c \, \Phi 
\end{eqnarray}
These terms arise after the boundary Higgs sector has developed its
VEV structure.
As commented in Ref.~\cite{Carone:2005rk}, these terms remain at the electroweak order as $\chi / \Lambda \sim \mathcal{O}(1)$ 
in the decoupling limit $v_i \rightarrow \Lambda \rightarrow \infty$, with their interactions dictated only by 
$G_{SM}\otimes SU(2)_{\ell}$ not the UV physics. 

These couplings return electroweak scale Dirac masses to the SM quarks 
and GUT scale masses to the exotic quarks, with mixing between the $d$ and $h$ 
quarks heavily suppressed. 
Analysis of the leptonic sector reveals that the 
electron does not mix at all with the exotic particles, 
and the mixing between the neutrinos and exotic neutral singlets 
is significantly suppressed. 
There are GUT scale Dirac masses for all the exotic particles, while 
the electron and neutrino acquire electroweak scale Dirac masses. We
shall not concern ourselves with naturally generating see-saw
suppressed Majorana masses for the neutrinos. We suspect that the
inclusion of gauge singlet fermions with Majorana masses at the
unification scale will invoke a seesaw style suppression of the
electroweak scale Dirac masses, as occurs in 4d
models~\cite{Demaria:2006uu}. However we shall settle for highly tuned Dirac
Yukawa coupling constants in what follows.
Nevertheless, the Yukawa interactions that are induced on the brane by
the Wilson line operators yield an appealing fermion mass
spectrum. We note that as our construct is non-supersymmetric
the usual fine tuning is required to stabilise the electroweak scale
relative to the unification scale.
\section{Gauge Coupling Unification\label{sec:quart_unification}}
We turn now to calculating the unification scale of our scheme. 
To evaluate the evolution of the gauge coupling constants, 
we need to ascertain how exactly the presence of 
the extra dimension affects the RGEs. 
We essentially have two mass scales to consider; 
the compactification scale which characterises the size of the extra dimension, and 
the cut-off or unification scale $\Lambda$.  

Corrections due to the physics in the UV regime arise in the form of brane localised operators. 
These respect the $G_{SM}\otimes SU(2)_\ell$ symmetry, and even if set to zero at 
tree-level can be generated by radiative corrections.
The effects of these operators, however, can be tamed, and much effort has been expended 
on ensuring that their presence does not destroy the higher dimensional unification \cite{gauge1,gauge2,gauge3}. 
The zero mode coupling in the effective 4d theory is obtained by integrating the 5d action for the gauge fields, 
and this estimates the value at the cut-off to be 
\begin{equation}\label{eq:couplingcorrect}
\centering
\frac{1}{g_{i}^2(\Lambda)}=\frac{2\,\pi\,R}{g_{5}^2(\Lambda)}+\frac{1}{\tilde{g}_{4\,i}^2(\Lambda)}.
\end{equation}
$ g_5$ is the $SU(3)^4$-invariant coupling parameter while $g_{\tilde{4}\,i}$ is the dimensionless coupling 
constant arising from these brane localised operators. 
If the volume factor of the extra dimension is sufficiently large then the sensitivity to 
these brane corrections is diluted by the bulk contribution \cite{gauge1}. 
If 
$g_5^2=\zeta/\Lambda, \, \tilde{g}_4^2=\zeta_a$ then the suppression is given by 
$(\zeta/\zeta_a)(1/\Lambda \pi R)$. So as long as the extra dimension is sufficiently large in 
extent, the volume factor $\Lambda \pi R$ is sufficient to negate the brane modification of $g_i(\Lambda)$.

What remains is to determine 
the threshold contributions which arise at the compactification scale $M_c$. 
At energies below $M_c$, the extra dimension is not observable. The zero-modes 
of the KK gauge boson towers and brane localised fermions define our effective theory to be that of the SM 
supplemented with the 
additional $SU(2)_\ell$ symmetry. Within this regime, the evolution of the coupling constants proceeds 
via the usual renormalisation group equations 
\begin{equation}
\centering
\frac{1}{\alpha_i(M_c)}=\frac{1}{\alpha_i(M_{EW})} + \frac{b_i}{2\,\pi}\,\ln{\frac{M_{EW}}{M_c}}.
\end{equation}
$\alpha_i = g_i^2/4\pi$ and the $b_i$ are the beta coefficients which enumerate both the number and type 
of particles which contribute to the evolution. 

Above $M_c$ the extra dimension is manifest through the appearance of the infinite towers of KK modes 
with increasing mass, and our effective theory is the 5-dimensional GUT with symmetry group $SU(3)^4$. The towers, however, do 
not have a universal contribution to the running of the coupling constants. This is in part because they do not fill complete 
$SU(3)^4$ representations, and also due to the misalignment of the towers resulting from the brane breaking. 
At $M_c$ the $G=SU(3)_q\otimes SU(2)_L \otimes SU(3)_\ell \otimes SU(3)_R \otimes U(1)$ symmetry is valid, with the 
modes corresponding to the $G/H$ fields starting to appear. These states contribute at each $(2\,n+1)\,M_c $ level. As 
the evolution proceeds from $M_c$ to $2M_c$ the full quartification symmetry emerges with the 
$SU(3)^4/G$ 
states inputting at the $(2n+2)M_c$ energy levels. At each $n$-th level, new excitation modes contribute until the couplings 
merge at $M_{GUT}$. 
It has been shown that if the difference on the runnings $\delta_i(\mu) \equiv  \alpha_i^{-1}(\mu)-\alpha_j^{-1}(\mu)$ is 
considered then these KK modes dominate the evolution above $M_c$, with the differential running 
logarithmically sensitive to the cut-off as per \cite{gauge1,gauge2,gauge3}
\begin{equation}
\centering
\delta_i(M_{GUT}) =\delta_i(M_{EW}) +\frac{(\beta_i-\beta_j)}{2\,\pi}\,\ln{\frac{M_{EW}}{M_c}}-\frac{1}{2\,\pi}\,\Delta_i(M_{GUT}),
\end{equation}
where 
\begin{eqnarray}
\centering
\Delta_i(M_{GUT})&=&  (\beta_i-\beta_j) \, \ln{\frac{M_{GUT}}{M_c}} + 
(\gamma_i-\gamma_j) \, \sum_{n=0}^{N_k} \ln{\frac{M_{GUT}}{(2 \,n+1)\, M_c}}\nonumber \\
&+&(\eta_i-\eta_j) \, \sum_{n=0}^{N_k}\ln{\frac{M_{GUT}}{(2n+2)\,M_c}}.
\end{eqnarray}
Here, $\beta_i$ are the zero-mode beta coefficients, $\gamma_i$ are the beta coefficients of the modes with mass $(2n+1)M_c$ 
and $\eta_i$ those of the modes with mass $(2n+2)M_c$. 
These factors compensate for the varying contributions of the KK levels. 
$N_k$ is 
the level at which the KK towers are truncated, with $(2 N_k+1)M_c \lesssim M_{GUT}$. 
We choose to explore the running with respect to the evolution of $\alpha_Y$, taking $j=Y$. 

Table~\ref{tab:KKbetas} lists the beta coefficients for all the KK modes, 
where the gauge bosons have been re-labelled in terms of their $G_{SM} \otimes SU(2)_\ell$ representations. 
The multiplet $A_{SM'}$ consists of the $G_{SM} \otimes SU(2)_\ell$ gauge fields, while the EW Higgs is contained in 
$A_{(\mathbf{1,2,1})(\pm 1)}$. The remaining two multiplets represent exotic fields contained in $G/H$. 
These arise from the $A^\mu_R$ fields aligned along the $T^4, \, T^5$ directions and those corresponding to 
$A_R^{\mu \, a=1,2}$ whose KK masses have received a shift. 

\begin{figure}
\begin{center}
\scalebox{0.5}{\includegraphics{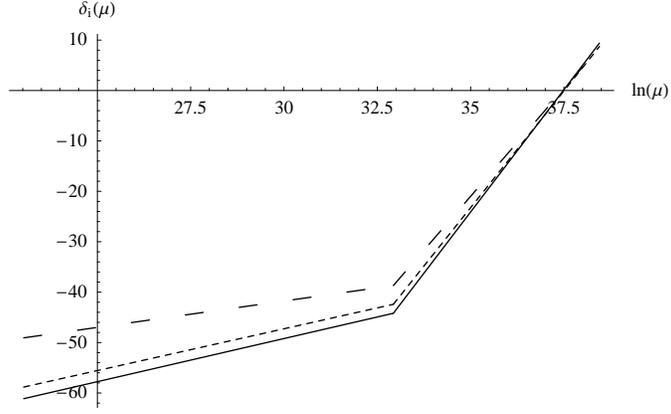}}
\caption{The differential running of the SM gauge coupling constants for a compactification scale of $M_c \sim 2 \times 10^{14}GeV$ 
as a function of $\ln{(\mu)}$. The solid line is
$\delta_{\ell}$, the short-dashed line is $\delta_q$, the
long-dashed line is $\delta_L$ and the condition for unification is
$\delta_\ell=\delta_q=\delta_L=0$. This occurs at 
$M_{GUT} \sim 1.9 \times 10^{16} \, GeV$.}
\label{fig:orbrge3}
\end{center}
\end{figure}

Given that the exotic states begin to emerge at $M_c$, the compactification scale has the lower bound 
$M_c \gtrsim 1 \,TeV$. 
If unification could result at this compactification energy, then there would be 
a promising spectrum of exotic particles within reach of the LHC. 
Unfortunately the three SM coupling 
parameters do not unify with a low $M_c$. The lowest value of the compactification scale which is consistent with approximate 
unification is $M_c \sim 10^{10} \, GeV$, with unification occuring at
$M_{GUT}\sim 10^{12} \,GeV$. As we increase the compactification scale
further, the energy interval between $M_c$ and $M_{GUT}$ decreases slightly, 
and fewer KK modes contribute to the running. The most favourable unification scenario occurs for $M_c \sim 2 \times 10^{14}\,GeV$. 
Here we have $\sim 50$ KK states for each tower in the summation with the coupling constants intersecting at 
$M_{GUT} \sim 1.9 \times 10^{16} \, GeV$ as shown in
Fig.~\ref{fig:orbrge3}. In this case, the leptonic colour coupling parameter must be
$\alpha_\ell \sim 0.19$ at the electroweak level, surpassing the
electroweak scale value of the strong coupling constant. We reiterate that the differential running has
been defined relative to the hypercharge coupling constant so that
\begin{eqnarray}
\delta_{q, L, \ell}(\mu)=\alpha^{-1}_{q, L,
  \ell}(\mu)-\alpha^{-1}_Y(\,u).
\end{eqnarray}
To complete the unification analysis we determine the largest possible compactification scale which allows for 
unification. As $M_c$ increases in energy, necesarily so too does the unification scale. Thus the upper bound on 
$M_c$ will result when the GUT scale is identified as the 5d Planck scale. 
This is attained for $M_c \sim 10^{16}\,GeV$ which gives unification
at $M_{GUT} \sim 8.2 \times 10^{17}\,GeV$. To summarise we find that
the relation $\delta_i(M_{GUT})=0, \,\forall \,i=q,L,\ell,Y$ 
is satisfied only for a compactification scale of order $10^{10} \,GeV
< M_c < 10^{16}\, GeV$, with the corresponding unification scale
lying in the range
$10^{12} \, GeV < M_{GUT} < 8 \times 10^{17}\, GeV$. Within these bounds, the 
amalgamation of the strong and electroweak couplings constrains the value 
of the leptonic colour coupling parameter to be $0.08 \lesssim \alpha_\ell(M_{EW}) \lesssim 0.25 $ at the electroweak 
scale.

\begin{table}
\begin{center}
\begin{tabular}{ccc}
\hline
\hline
Multiplet & $(b_Y,b_L,b_q,b_\ell)$ & $(\tilde{b}_Y,\tilde{b}_L,\tilde{b}_q,\tilde{b}_\ell)$ \\
\hline
\hline
$A_{SM'}$ & $\left(0,-\frac{22}{3},-11,-\frac{22}{3}\right)$ & $\left(0,-\frac{20}{3},-10,-\frac{20}{3}\right)$ \\
$A_{(\mathbf{1,2,1})(- 1)}$ & $\left(\frac{1}{12},\frac{1}{6},0,0\right)$ &$\left(-\frac{5}{6},-\frac{7}{2},0,0\right)$\\
\hline 
$(\eta_Y,\eta_L,\eta_q,\eta_\ell)$ & &$\left(-\frac{5}{6},-\frac{61}{6},-10,-\frac{20}{3}\right)$ \\
\hline
$A_{(\mathbf{1,1,2})(-1)}$ & ------ & $\left(-\frac{5}{6},0,0,-\frac{7}{2}\right)$\\
$A_{(\mathbf{1,1,1})(\pm2)}$ & ------ & $\left(-\frac{10}{3},0,0,0\right)$\\
\hline
$(\gamma_Y,\gamma_L,\gamma_q,\gamma_\ell)$ & &$\left(-\frac{25}{6},0,0,-\frac{7}{2}\right)$\\
\hline
matter & $\left(\frac{10}{3},4,4,0\right)$ & ------ \\
\hline
\hline
TOTAL & $\left(\frac{41}{12},-\frac{19}{6},-7,-\frac{22}{3}\right)$&$\left(-5,-\frac{61}{6},-10,-\frac{61}{6}\right)$\\
\hline
\hline
\end{tabular}
\caption{The enumeration of the KK modes and their contribution to the RGEs. The $b$'s represent the zero mode 
beta coefficients, while $\tilde{b}_i$ reflect the excitation modes. Here the decomposition of the 5d gauge fields 
is in terms of their $H$ representation.} 
\label{tab:KKbetas}
\end{center}
\end{table}

\section{A model without $SU(2)_{\ell}$\label{sec:quart_broken_su2}}
We now briefly comment on 5d quartified models which do not have a residual leptonic $SU(2)_{\ell}$ 
symmetry, i.e. $H=G_{SM}$. There exists only a limited approach to extend upon the previous scheme in which this symmetry 
remained in the low energy theory. We require the breaking of the leptonic symmetry to be achieved by a combination 
of the orbifolding and brane localised scalar fields. 

To obtain a consistent low-energy model the 
orbifold projections must be  
\begin{eqnarray}\label{eq:noSU2orbP}
\centering
P_q&=&\textrm{diag}(1,1,1), \;\;\;\; P_L=\textrm{diag}(1,1,-1),\;\;\;\;P_\ell=\textrm{diag}(1,1,1), \;\;\;\; P_R=\textrm{diag}(1,1,-1),\nonumber \\
P_q'&=&\textrm{diag}(1,1,1), \;\;\;\;P_L'=\textrm{diag}(1,1,-1),\;\;\;\; P_\ell'=\textrm{diag}(1,1,-1), \;\;\;\; P_R'=\textrm{diag}(1,1,1),
\end{eqnarray}
which break the quartification symmetry to $SU(3)_q \otimes SU(2)_L \otimes SU(2)_{\ell} \otimes SU(3)_R \otimes U(1)_L\otimes U(1)_{\ell}$ 
on the matter brane. 
As the quarks are singlets under leptonic colour, we still require the full $SU(3)_R$ symmetry on the brane and the fields 
$\chi_1$ and $\chi_2$ to generate mass terms. Hence, our quark sector will be identical to that of the previous model. 
To instigate the breaking to the SM gauge group we need also the additional 
boundary scalar fields 
\begin{equation}\label{eq:newscalars}
\centering
\chi_{3}  \sim ({\mathbf{1,1,1,3}})(0,-2/\sqrt{3}), \qquad \chi_4 \sim ({\mathbf{1,1,2,1}})(-2/\sqrt{3},-1/\sqrt{3}), \qquad \chi_5  \sim ({\mathbf{1,1,2,3}})(0,1/\sqrt{3}),
\end{equation}
with the VEV patterns 
\begin{eqnarray}
\centering
\langle \chi_3 \rangle &=& \left( \begin{array}{c} 0 \\ 0 \\ v_4 \end{array} \right), \qquad 
\langle \chi_4 \rangle =  \left( \begin{array}{c} v_5 \\ 0 \end{array} \right) \nonumber \\
\langle \chi_5 \rangle &=& \left( \begin{array}{ccc} 
v_6 & 0 & v_7 \\
0 & v_8 & 0 \end{array} \right).
\end{eqnarray}

We now have only four charged lepton fields per family and they all mix. The Yukawa interaction terms generated locally 
through $\chi_1-\chi_4$ and those induced non-locally with $\Phi$ impart an electroweak scale 
Dirac mass to the electron, muon and tauon particles. The remaining three eigenvalues per family are all of GUT scale. 
The neutral particles again have Dirac masses, 
with four of the eigenvalues at the GUT scale, and the other of EW order
which we identify with the neutrino. 
Thus the brane localised Higgs fields 
break the symmetry down to the SM gauge group and deliver mass terms to the exotic particles. At the 
electroweak scale, the theory comprises of the minimal SM particle spectrum and the right-handed neutrino. 
The exotic fermions all have heavy masses, and are decoupled from the theory in the Higgsless limit. 

Above the compactification scale, the excitation modes of the SM gauge fields and the multiplet containing the 
SM Higgs field appear at each $(2n+2)M_c$ level. Four exotic singlets 
with hypercharge $Y=\pm 2$ also emerge at the $(2n+1)M_c$ energy levels, with three of these arising from the shifted 
$(+,+)$ towers of $A^\mu_\ell$ and $A^\mu_R$. 

We have investigated the prospects for unification with this new
symmetry framework. We find that the fewer states contributing at each
$(2n+1)M_c$ level fail to sway the coupling constants to unify in a
phenomenologically consistent fashion. The couplings converge only 
when $M_c \ge 10^{16} GeV$. However, for compactification scales this
large, the unification always occurs at energies 
greater than the 5d Planck scale and so cannot result. It is
interesting that unification within our 5d framework requires
$SU(2)_\ell$ to remain as an exact symmetry.

\section{Camparison With Conventional
  Models\label{sec:quart_comparison}}
We have shown that unification can be achieved within a five dimensional
quartification model. Before concluding we shall contrast the 5d
construct with the conventional quartification models of ~\cite{Demaria:2005gk,Demaria:2006uu}. Our
intention is to draw the readers attention to the advantages and
disadvantages of each approach.

Let us first consider the case of unbroken $SU(2)_\ell$. In the 4d
case unification could be achieved with four distinct models, each
differing in their symmetry breaking pattern. No predictions regarding
the favourability of the distinct models was possible with all of
these models permitting a flexible range of symmetry breaking scales
consistent with unification. However
intermediate symmetry breaking scales were required, otherwise
unification mandates the inclusion of 14-37 SM Higgs doublets. 
The models with intermediate breaking scales required eight distinct
scalar $SU(3)^4$
multiplets (related by the cyclic $Z_4$ symmetry), and these  
naturally returned 
seven SM Higgs doublets. 
Furthermore, this results in complicated Higgs potentials with many
arbitrary parameters. These parameters must be tuned to generate the
desired stages of symmetry breaking, with further assumptions required
to ensure that the masses of the scalars were consistent with
unification. The GUT scale was found to lie in the range
$10^{12}-10^{18}$~GeV and proton decay inducing scalars were required
to be of order the GUT scale to ensure proton longevity. Finally 
we note that several of the 4d models predicted that the $SU(2)_R$ $W$ and $Z'$
bosons should be order TeV and are thus observable at the LHC.

We contrast this with the 5d case. Here the demands of a realistic SM
fermion spectrum restricted the intermediate symmetry group
on the matter brane $G_{\pi R/2}$, resulting in just one feasible choice, namely $G_{\pi
  R/2} \equiv SU(3)_q \otimes SU(2)_L \otimes SU(3)_\ell \otimes
SU(3)_R \otimes U(1)_L$. We have shown that unification can be
achieved via this route, occurring for a range of compactification
scales, $10^{10}\lesssim M_c\lesssim 10^{16}$~GeV. Only one SM Higgs
doublet arose in the 5d construct and this was adequate to achieve
unification. The reduced symmetry operative on the matter brane meant
that brane localised scalars need not fill out entire $SU(3)^4$
representations. Thus no coloured scalars were required and the proton
was found to be stable. This has the advantage of permitting
unification at low scales (relative to typical 4d unification
scales). The 5d framework naturally motivates the intermediate mass
scales necessary for unification. These are introduced in a somewhat
ad hoc way in 4d constructs, but here they arise as KK excitations of
bulk fields. Thus all intermediate mass scales are set by the one
parameter, $M_c$, and the inclusion of intermediate mass scales
doesn't imply an increase in parameter numbers. As we consider the
Higgsless limit the complications which arise in the Higgs sector of
4d models essentially disappear. However, no new phenomenology appears
in our model until the scale $M_c$ and given the large $M_c$ values required
to achieve unification this prohibits the observation of any new particles
at the LHC. This is similar to the 5d trinification models of
~\cite{Carone:2005rk} which
reproduce the minimal supersymmetric SM at low energies but predict no
additional phenomenology.

The 4d models with broken $SU(2)_\ell$ found to unify in~\cite{Demaria:2005gk}
possessed essentially the same features as those outlined above for
the unbroken $SU(2)_\ell$ case and thus provide no real
advantage over these models. Given that unification cannot be achieved
without a remnant $SU(2)_\ell$ symmetry in our 5d model we shall not comment further
on these.
\section{Conclusion\label{sec:quart_conc}}
We have investigated five dimensional quartification models where the
symmetry breaking is achieved by a combination of orbifold
compactification and the introduction of a boundary Higgs sector. We
have shown that the
SM Higgs doublet may be identified as the fifth component of a higher
dimensional gauge field. This forces matter to be brane localised, with
the SM Yukawa structure arising from fermion couplings with Wilson
line operators. The models may be considered in a Higgsless limit
wherein all gauge fields corresponding to generators broken above the
electroweak scale have their mass set by the compactification
scale. As in 4d models, intermediate mass scales were required to
ensure unification. However only
one arbitrary scale, namely the compactification
scale, was required, with the embedding of the quartification model in
a higher dimensional framework
naturally introducing a new class of threshold corrections to the running
coupling constants, corresponding to KK excitations of bulk gauge fields.  Surprisingly we found that
a unique symmetry breaking pattern consistent with both unification
and phenomenological demands
emerged from our framework. This required $SU(2)_\ell$ to remain as an
exact low-energy symmetry and differed markedly from the 4d case where
multiple symmetry breaking routes consistent with unification have
been found for both broken and unbroken $SU(2)_\ell$
models. Importantly the higher dimensional model alleviates the
complications arising from the Higgs sector of 4d models by rendering
this sector supplementary.
\section*{Acknowledgements}
This work was
supported in part by the Australian Research Council and in part by the Commonwealth of Australia. The authors thank R. R. Volkas and
C. D. Carone for helpful communications.


\begin{thebibliography}{99}
\bibitem{Foot:1990dw}
  R.~Foot and H.~Lew,
  Phys.\ Rev.\ D {\bf 41} (1990) 3502.
\bibitem{Foot:1990um}
  R.~Foot and H.~Lew,
  Phys.\ Rev.\ D {\bf 42} (1990) 945.
\bibitem{Foot:1990un}
  R.~Foot and H.~Lew,
  Mod.\ Phys.\ Lett.\ A {\bf 5}, 1345 (1990).
\bibitem{Foot:1991fk}
  R.~Foot, H.~Lew and R.~R.~Volkas,
  Phys.\ Rev.\ D {\bf 44}, 1531 (1991).
\bibitem{Levin:1993sq}
  Y.~Levin and R.~R.~Volkas,
  Phys.\ Rev.\ D {\bf 48}, 5342 (1993)
  [arXiv:hep-ph/9308256].
\bibitem{Shaw:1994zs}
  D.~S.~Shaw and R.~R.~Volkas,
  Phys.\ Rev.\ D {\bf 51}, 6490 (1995)
  [arXiv:hep-ph/9410350].
\bibitem{Foot:1995xx}
  R.~Foot and R.~R.~Volkas,
  Phys.\ Lett.\ B {\bf 358}, 318 (1995)
  [arXiv:hep-ph/9505331].
\bibitem{McDonald:2006dy}
  K.~L.~McDonald and B.~H.~J.~McKellar,
  Phys.\ Rev.\ D {\bf 74}, 056005 (2006)
  [arXiv:hep-ph/0609110].
\bibitem{Joshi:1991yn}
  G.~C.~Joshi and R.~R.~Volkas,
  Phys.\ Rev.\ D {\bf 45}, 1711 (1992).
\bibitem{Babu:2003nw}
  K.~S.~Babu, E.~Ma and S.~Willenbrock,
  Phys.\ Rev.\ D {\bf 69}, 051301 (2004)
  [arXiv:hep-ph/0307380].
\bibitem{Glashow:1984gc}
  S.~L.~Glashow,
Print-84-0577 (BOSTON)
\bibitem{Babu:1985gi}
  K.~S.~Babu, X.~G.~He and S.~Pakvasa,
  Phys.\ Rev.\ D {\bf 33}, 763 (1986).
\bibitem{Kim:2004pe}
  J.~E.~Kim,
  Phys.\ Lett.\ B {\bf 591}, 119 (2004) [arXiv:hep-ph/0403196].


\bibitem{Carone:2005rk}
  C.~D.~Carone and J.~M.~Conroy,
  Phys.\ Lett.\ B {\bf 626}, 195 (2005)
  [arXiv:hep-ph/0507292].
\bibitem{Carone:2004rp}
  C.~D.~Carone and J.~M.~Conroy,
  Phys.\ Rev.\ D {\bf 70}, 075013 (2004)
  [arXiv:hep-ph/0407116].
\bibitem{Demaria:2006uu}
  A.~Demaria, C.~I.~Low and R.~R.~Volkas,
  Phys.\ Rev.\ D {\bf 74}, 033005 (2006)
  [arXiv:hep-ph/0603152].
\bibitem{Demaria:2005gk}
  A.~Demaria, C.~I.~Low and R.~R.~Volkas,
  Phys.\ Rev.\ D {\bf 72}, 075007 (2005)
  [Erratum-ibid.\ D {\bf 73}, 079902 (2006)]
  [arXiv:hep-ph/0508160].
\bibitem{Kim:2002im}
 H.~D.~Kim and S.~Raby,
  JHEP {\bf 0301}, 056 (2003) [arXiv:hep-ph/0212348].
\bibitem{gaugehiggs}  
L.~J.~Hall, Y.~Nomura and D.~R.~Smith,
  Nucl.\ Phys.\ B {\bf 639}, 307 (2002)  [arXiv:hep-ph/0107331];
  G.~Burdman and Y.~Nomura,
Nucl.\ Phys.\ B {\bf 656}, 3 (2003) [arXiv:hep-ph/0210257];
 N.~Haba, Y.~Hosotani, Y.~Kawamura and T.~Yamashita,  
  Phys.\ Rev.\ D {\bf 70}, 015010 (2004) [arXiv:hep-ph/0401183];
  C.~Csaki, C.~Grojean and H.~Murayama,
  
Phys.\ Rev.\ D {\bf 67}, 085012 (2003)  [arXiv:hep-ph/0210133];  
  C.~A.~Scrucca, M.~Serone and L.~Silvestrini,
  
Nucl.\ Phys.\ B {\bf 669}, 128 (2003)  [arXiv:hep-ph/0304220];
  N.~Haba and Y.~Shimizu,   
Phys.\ Rev.\ D {\bf 67}, 095001 (2003)  [Erratum-ibid.\ D {\bf 69}, 059902 (2004)]  [arXiv:hep-ph/0212166];
 K.~W.~Choi, N.~Y.~Haba, K.~S.~Jeong, K.~i.~Okumura, Y.~Shimizu and M.~Yamaguchi,
  
JHEP {\bf 0402}, 037 (2004)  [arXiv:hep-ph/0312178];
I.~Gogoladze, Y.~Mimura and S.~Nandi,  
  Phys.\ Lett.\ B {\bf 562}, 307 (2003)  [arXiv:hep-ph/0302176];
Y.~Hosotani,
  Phys.\ Lett.\ B {\bf 126}, 309 (1983);
 Y.~Hosotani,  
  
Annals Phys.\  {\bf 190}, 233 (1989);
  N.~Arkani-Hamed, L.~J.~Hall, Y.~Nomura, D.~R.~Smith and N.~Weiner,
  
Nucl.\ Phys.\ B {\bf 605}, 81 (2001)  [arXiv:hep-ph/0102090];  
  A.~Masiero, C.~A.~Scrucca, M.~Serone and L.~Silvestrini,
  
Phys.\ Rev.\ Lett.\  {\bf 87}, 251601 (2001)  [arXiv:hep-ph/0107201].
\bibitem{orbifolds}
 I.~Antoniadis, 
 Phys.\ Lett.\ B {\bf 246}, 377 (1990);
 I.~Antoniadis and K.~Benakli, Phys.\ Lett.\ B {\bf 326}, 69 (1994); 
 K.~Akama,
  Lect.\ Notes Phys.\  {\bf 176}, 267 (1982); 
 Y.~Kawamura,
  Prog.\ Theor.\ Phys.\  {\bf 103}, 613 (2000)  [arXiv:hep-ph/9902423];
  Y.~Kawamura,  
  
Prog.\ Theor.\ Phys.\  {\bf 105}, 999 (2001)  [arXiv:hep-ph/0012125];  
  Y.~Kawamura,
  
Prog.\ Theor.\ Phys.\  {\bf 105}, 691 (2001)  [arXiv:hep-ph/0012352]; 
 G.~Altarelli and F.~Feruglio,
A.~Hebecker and J.~March-Russell,  
  
Nucl.\ Phys.\ B {\bf 613}, 3 (2001) [arXiv:hep-ph/0106166].

\bibitem{Csaki:2003dt}
  C.~Csaki, C.~Grojean, H.~Murayama, L.~Pilo and J.~Terning,
Phys.\ Rev.\ D {\bf 69}, 055006 (2004)  [arXiv:hep-ph/0305237].
\bibitem{Coulthurst:2006bc}
  A.~Coulthurst, K.~L.~McDonald and B.~H.~J.~McKellar,
  Phys.\ Rev.\ D {\bf 74}, 127701 (2006)
  [arXiv:hep-ph/0610345].
\bibitem{gauge1}
 Y.~Nomura 
Phys.\ Rev.\ D {\bf 65} 085036, (2002);
\bibitem{gauge2}
Hall, L. J. and Nomura, Y., Phys.\ Rev.\ D {\bf 65} 125012, (2002).
\bibitem{gauge3}
Phys.\ Lett.\ B {\bf 511}, 257 (2001) [arXiv:hep-ph/0102301];
L.~J.~Hall and Y.~Nomura, 
 Phys.\ Rev.\ D {\bf 64}, 055003 (2001)  [arXiv:hep-ph/0103125];
Y.~Nomura, D.~R.~Smith and N.~Weiner,  
  
Nucl.\ Phys.\ B {\bf 613}, 147 (2001)  [arXiv:hep-ph/0104041].




\end{thebibliography}
\end{document}